\documentclass[conference]{IEEEtran}

\usepackage{amsmath,amssymb,amsfonts,bm}
\usepackage{algorithm,algorithmic}
\usepackage{graphicx} 

\ifCLASSINFOpdf

\else

\fi

\DeclareMathOperator*{\argmax}{arg\,max}

\begin{document}

\title{Centralized QoS-Aware Resource Allocation for D2D Communications with Multiple D2D Pairs in One Resource Block}

\author{Mohammad Hossein Bahonar$^\dagger$, Mohammad Javad Omidi$^\dagger$\\
$^\dagger$ Dept. of Electrical and Computer Engineering, Isfahan University of Technology, Isfahan, Iran 84156-83111\\
Emails: \{mh.bahonar@ec.iut.ac.ir, omidi@cc.iut.ac.ir\}
}

\maketitle

\begin{abstract}
Device-to-device (D2D) communications can result in spectral efficiency (SE) improvement of cellular networks when interference is addressed
 properly.
For further improvement of overall network sum-rate, it can be assumed that multiple D2D pairs reuse the same resource block (RB) of a cellular user equipment (CUE).
In this paper, we study a centralized joint power and spectrum resource allocation problem to maximize the overall sum-rate of the network while guaranteeing the quality-of-service (QoS) requirements of CUEs and admitted D2D pairs.
The proposed method uses a distance based approach to select a CUE at each step with a distance and power based approach to increase the number of admitted D2D pairs.
A power minimization approach when the number of admitted D2D pairs are increasing and a power maximization approach at the end of resource allocation for each RB is proposed.
Numerical results show that the proposed algorithm significantly improves SE compared to the situation where at most one D2D pair can reuse the RB of a CUE.
It also shows that the proposed algorithm has a slight lower performance compared to the optimal scheme but dramatically decreases the matching complexity and signaling overhead.
\end{abstract}

\IEEEpeerreviewmaketitle

\section{Introduction}
Due to the ever increasing data demand, a device-to-device (D2D) communication scheme has been proposed which enables direct data transmission between two devices in proximity without going through the base station (BS).
This type of communication can increase the overall network sum-rate and spectral efficiency (SE). 
Power control \cite{R052}, security, interference management \cite{R011}, caching, coverage analysis \cite{R053}, and resource allocation \cite{R003} are some of the challenges in D2D communications.

Interference management is an important issue in D2D communications since this type of communication causes interference to existing cellular user equipment (CUE) in underlaying schemes.
The resource allocation problem can be formulated as a constraint optimization problem with SE \cite{R024_061} or energy efficiency (EE) \cite{R040} objective functions and quality-of-service (QoS) constraints in terms of signal-to-interference-plus-noise ratio (SINR) or achievable data rate in an uplink \cite{R050} or downlink resource reuse \cite{R024} schemes.
In \cite{R003} a 3-step optimal centralized resource allocation algorithm is proposed that assumes at most one D2D pair in each resource block (RB).
It is usually assumed that the BS has complete channel state information (CSI) of all links \cite{R004}, in the centralized scheme.
In \cite{R024_061} channel uncertainty is investigated and resources are allocated in a centralized manner.
Due to the high complexity of centralized approaches, distributed approaches \cite{R043} are used with lower complexity and signaling overhead that have some performance losses compared to the centralized schemes.
The QoS constraints of the optimization algorithm limit the interference experienced by the CUEs.
The QoS constraints can be either related to just the CUEs since they are considered as primary users or to both CUEs and D2D pairs \cite{R003} since the achievable data rate requirement of D2D pairs is also of high importance.
Consideration of QoS for D2D pairs makes the optimization problem more complicated, which makes researchers to focus more on the QoS constraints of CUEs \cite{R045}.

Each CUE occupies one RB in a cellular network.
It is usually assumed that at most one D2D pair can reuse one RB, while consideration of multiple D2D pairs in one RB can result in significant performance improvement of the network.
Consideration of multiple D2D pairs in one RB results in a dramatic increase of complexity and a signaling overhead.
Due to the high complexity and  signaling overhead of centralized schemes, it is not reasonable to consider multiple D2D pairs in centralized scheme without complexity and signaling overhead reduction.
In \cite{R011}, multiple D2D pairs in one RB are assumed and the optimization problem is solved using a distributed approach that has a lower performance compared to the centralized ones.
The proposed method allocates resource to each UE based on its own CSI which results in low complexity and signaling overhead of the proposed method.
It is assumed that the number of D2D pairs is more than the number of CUEs in \cite{R041} and network capacity optimization is considered.
Resource allocation with the assumption of multiple D2D pairs in one RB is investigated in \cite{R042_026}, but the QoS of CUEs is just guaranteed.
The resource allocation problem is modeled as a mixed integer programming optimization problem and solved using a subchannel sharing protocol.
Multiple D2D pairs in one RB assumption is also considered in \cite{R050}, while the QoS of CUEs is only considered and the EE of D2D links is optimized.

We formulate the resource allocation problem to allocate power and spectrum to CUEs and D2D pairs in a cellular network in order to maximize the overall sum-rate while the D2D pairs reuse the uplink spectrum.
The problem is modeled as a constraint optimization problem, where the QoS of CUEs and D2D pairs in terms of SINR are considered as the optimization constraints.
We propose selecting one CUE at each step based on a distance metric and allocate the maximum number of D2D pairs to the RB of the CUE while the optimization problem remains feasible.
The procedure of adding D2D pairs to the RB of the selected CUE is performed at several steps.
At each step a D2D pair is selected based on a distance and power metric. Then, the power of CUEs using the same RB is minimized in order to add more D2D pairs to the RB, that results in further sum-rate increment.
At the end of the resource allocation to each RB, power of CUEs using the same RB is optimized to increase the overall sum-rate.
It is shown that the proposed scheme has a slight lower performance compared to the optimal scheme but it dramatically decreases the matching complexity and signaling overhead of resource allocation.

The rest of the paper is organized as follows.
System model consisting of the network model and the achievable data rate alongside with the notation of used symbols are introduced in section \ref{SecSymModel}.
Problem formulation for power and spectrum allocation to CUEs and D2D pairs with multiple D2D pairs in one RB is presented at section \ref{SecFormulation}.
The proposed centralized resource allocation algorithm is given in section \ref{SecResourceAlloc}.
Complexity and signaling overhead is analyzed in section \ref{SecComp}.
Numerical results are presented at section \ref{SecNumeric} and section \ref{SecConclusion} concludes the paper.

\section{System Model}
\label{SecSymModel}

The system bandwidth is divided into N orthogonal RBs.
Considering a fully loaded cellular scenario, $N$ active CUEs denoted by $C_1, C_2, ..., C_N$ occupy the $N$ orthogonal channels in the cell and $M$ D2D pairs denoted by $D_1, D_2, ..., D_M$ coexist with CUEs.
We use $\textbf{c}=\{1, ...,N \}$ and $\textbf{d}=\{1, ..., M \}$ to denote the index sets of CUEs and D2D pairs.
Each RB is allocated to one CUE and each CUE occupies one RB different from other CUEs.
Each D2D pair shares one RB with a CUE but one RB can be allocated to multiple D2D pairs.
Uplink resource sharing is considered. 
Since the BS is located  at the center of the cell and has a higher transmission power compared to other UEs, downlink interference is mostly caused by the BS.
This kind of interference exists in large areas of the cell while uplink interference exists just around the CUE that uses its corresponding RB.
Each D2D transmitter sends its data toward its corresponding receiver in the uplink RB of a CUE.
Based on the value of $M$, the number of D2D pairs might be much larger than the number of available RBs.
So it is necessary for multiple D2D pairs to occupy the RB of one CUE simultaneously.
Also each CUE and D2D pair have their own minimum QoS requirement in terms of SINR.

The channel gain between CUE $i$ and the BS is denoted as $g_{C_i,B}$ and that of the D2D pair $i$ between its transmitter and receiver is denoted by $g_{D_i}$.
The channel gain from the transmitter of D2D pair $i$ to the BS, from the CUE $i$ to receiver of D2D pair $j$, and from transmitter of D2D pair $i$ to receiver of D2D pair $j$ are denoted as $h_{D_i,B}$, $h_{C_i,D_j}$, and $h_{D_i,D_j}$, respectively.
The power of additive white Gaussian noise (AWGN) on each channel is assumed to be $\sigma_N^2$ while $\sigma_s^2$ denotes the power of signal processing noise at the receiver.
Both  fast fading due to multi-path propagation and slow fading due to shadowing have also been considered. 

Let $\gamma_{C_i}$ and $\gamma_{D_j}$ denote the SINR for the link between CUE $i$ and the BS and the link between transmitter and receiver of D2D pair $j$ respectively.
Also, let $\rho_{i,j}$ be the resource reuse indicator for CUE $i$ and D2D pair $j$, $\rho_{i,j}=1$ when D2D pair $j$ reuses the resource of cellular user $i$; otherwise, $\rho_{i,j}=0$.
Considering a total bandwidth of $W$, the bandwidth of each RB would be $B_{\rm RB}=\frac{W}{N}$. 
The SINR of CUE $i$ can be expressed as
\begin{equation}
\label{EqGammaC}
\gamma_{C_i} = \frac{P_{C_i} g_{C_i,B}}{ \sum_{j} \rho_{i,j} P_{D_j}h_{D_j,B} + \sigma_N^2 + \sigma_s^2}
\end{equation}
where $P_{C_i}$ denotes the transmission power of CUE $i$.
The summation at the denominator of \eqref{EqGammaC} corresponds to interference from other D2D pairs that use the same RB of CUE $i$.
The SINR of D2D pair $i$ can be expressed as
\begin{equation}
\label{EqGammaD}
\gamma_{D_i} = \frac{P_{D_i} g_{D_i}}{ 
\sum_{j} \rho_{j,i} [ P_{C_j}h_{C_j,D_i} + \sum_{k\neq i} \rho_{j,k} P_{D_k} h_{D_k,D_i}  ] 
+ \sigma_N^2 + \sigma_s^2}
\end{equation}
where $P_{D_i}$ denotes the transmission power of D2D pair $i$.
The summation at the denominator of \eqref{EqGammaD} corresponds to interference from one CUE and other D2D pairs using the same RB of D2D pair $i$.

\section{Problem Formulation}
\label{SecFormulation}
The purpose of this work it to maximize the sum-rate performance of all UEs in a cell. 
Since there exists $N$ CUEs and $M$ D2D pairs in a cell, the sum-rate of all UEs in a cell, $R_{\rm Total}$ can be expressed as
\begin{equation}
\label{EqRTotal}
R_{\rm Total} = \sum\limits_{i\in \textbf{c}}{\log_2{(1+\gamma_{C_i})}} + \sum\limits_{i\in \textbf{d}}{\log_2{(1+\gamma_{D_i})}} 
\end{equation}
where the first and second summations correspond to sum-rate of CUEs and D2D pairs, respectively.
A cellular link from a CUE to the BS or a D2D link between the transmitter and the receiver of a D2D pair is set up only when their minimum QoS requirement can be guaranteed.
So the sum-rate maximization problem is an optimization problem constrained to some QoS requirement constraints that can be expressed as
\begin{subequations}
\begin{alignat}{3}
\label{EqOpt11}
&\max_{\textbf{p}_C, \textbf{p}_D, \bm{\rho}} \quad && R_{\rm Total} \\
\label{EqOpt12}
&s.t. && \gamma_{C_i} \geq \gamma_{C_i,{\rm min}},  \forall i \in \textbf{c} \\
\label{EqOpt13}
& && \gamma_{D_j} \geq K_j \gamma_{D_j,{\rm min}},  \forall j \in \textbf{d} \\
\label{EqOpt14}
& && K_j = \sum\limits_{i\in \textbf{C}} {\rho_{i,j}} \leq 1, \forall j \in \textbf{d}, \rho_{i,j} \in \{0,1\},   \\
\label{EqOpt15}
& && 0 \leq P_{C_i} \leq P_{C,{\rm max}} ,  \forall i \in \textbf{c} \\
\label{EqOpt16}
& && 0 \leq P_{D_j} \leq P_{D,{\rm max}} ,  \forall j \in \textbf{d} 
\end{alignat}
\end{subequations}
where $\textbf{p}_C=\{P_{C_1}, ..., P_{C_N} \}$ and $\textbf{p}_D=\{P_{D_1}, ..., P_{D_N} \}$ are transmit power vectors related to CUEs and transmitters of D2D pairs respectively.
$\gamma_{C_i,{\rm min}}$ and $\gamma_{D_j,{\rm min}}$  correspond to the minimum required QoS of CUE $i$ and D2D pair $j$, respectively.
$K_j$ is a reuse indicator variable related to D2D pair $j$. 
When D2D pair $j$ reuses the RB of a CUE, $K_j=1$ and we call the D2D pair an admitted D2D pair.
When D2D pair $j$ can not reuse the RB of any CUE, $K_j=0$ and we call it a denied D2D pair.
Constraints \eqref{EqOpt12} and \eqref{EqOpt13}  represent the QoS requirements of CUEs and admitted D2D pairs.
According to constraint \eqref{EqOpt13}, the achievable QoS related to a denied D2D pair is equal to zero.
Constraint \eqref{EqOpt14} indicates that a D2D pair shares at most the RB of one CUE.
Constraints \eqref{EqOpt15} and \eqref{EqOpt16} guarantee  the transmit powers of CUEs and D2D pairs, respectively.

\section{Centralized Resource Allocation }
\label{SecResourceAlloc}
The sum-rate maximization problem \eqref{EqOpt11} is a nonlinear constraint joint optimization problem of power control and matching between D2D pairs and CUEs.
We jointly design the matching variable $\bm{\rho}$ and the power control strategies $\textbf{p}_C, \textbf{p}_D$ to maximize overall sum-rate of the network.
Considering more than one D2D pair reusing the RB of a CUE results in increased complexity of the matching process.
Also due to the centralized resource allocation scheme and the assumption of known channel gains at the BS, multiple D2D pairs in one RB result in increased signaling overhead of the network, which is because of transmitting measured channel gains to the BS.
We propose increasing the overall sum-rate of the network by increasing the number of admitted D2D pairs as a higher priority and maximizing the power of all CUEs as a lower priority, since increasing the power of CUEs results in logarithmic increase of its data rate.
Increasing the number of admitted D2D pairs can be done by limiting the transmission power of other UEs in the same RB that results in smaller interference areas.

\subsection{Matching Complexity and Signaling Overhead Reduction}
\label{SubSec1}
The solution of the optimization problem in \eqref{EqOpt11} consists of a CUE-D2D matching procedure and power optimization of the matched UEs in one RB.
High complexity matching process is due to the fact that the BS has to check every possible matching state to find the optimal CUE-D2D matching.
Our proposed method is a suboptimal algorithm with a very low complexity matching procedure that  chooses one CUE at each step based on a distance metric and matches the maximum number of denied D2D pairs with the chosen CUE based on a distance and power based metric while QoS requirements are guaranteed.

The BS allocates power to each UE based on its desired and interference channel gains that are computed at the UEs and transmitted to the BS.
In our system model where the number of D2D pairs is more than CUEs, a huge signaling overhead is needed to transmit the interference gains to the BS.
We propose to estimate and transmit the channel gains related to the communication links between UEs that are matched to the same RB at each step of the algorithm when they are needed for power allocation.

\subsection{First Admitted D2D Pair}
\label{SubSec2}
In order to solve the optimization problem based on matching the maximum possible number of D2D pairs to one specific UE, the objective function of the optimization problem can be expressed as 
$R_{Total} = \sum_{i\in \textbf{c}} \bigg ( {\log_2{(1+\gamma_{C_i})}
 +  \sum_{i\in \textbf{d}} {\rho_{i,j} \log_2{(1+\gamma_{D_i})}} } \bigg )$
where the overall sum-rate of the network, $R_{Total}$ is expressed as the summation of $R_i$s that corresponds to the sum-rate of all UEs using the RB of CUE $i$.

In the proposed algorithm, the CUEs with larger distance from the BS are selected earlier.
Let $s(A,B)$ denote the distance between A and B.
We define CUE priority vector $\bm{\alpha} = \{ \alpha_1, ... , \alpha_N \}$ as the vector of CUE indices that their distance from the BS has a descending order, i.e.,
\begin{equation}
\label{EqAlphaCompute}
s(C_{\alpha_1},BS) > ... > s(C_{\alpha_N},BS)
\end{equation}
where $s(C_{\alpha_i},BS)$ denotes the distance between $C_{\alpha_i}$ and the BS.
CUEs are selected based on the CUE priority vector.



After selecting the CUE $\alpha_i$, one denied D2D pair must be selected to reuse the RB of the CUE.
Since the distance highly influences the channel gain value and QoS constraints, we define $m_j$ equal to the distance of CUE $\alpha_i$ from the receiver of D2D pair $j$ as in \eqref{Eqmj}.
The selected denied D2D pair $j^*$ is the one with the largest distance from CUE $\alpha_i$.
\begin{align}
\label{Eqmj}
m_j &=s(C_{\alpha_i}, D_{j,Rx}) \\
\label{EqmjMax}
j^* &= \argmax_{j \in \mathcal{D}} {(m_j)}
\end{align}
where $\mathcal{D}$ is the set of denied D2D pairs.
The resource indicator variable is also updated, i.e., $ \rho_{\alpha_i,j^*} = 1$ .

In order to increase the overall sum-rate of the network by maximizing the number of admitted D2D pairs, it is necessary to lower the amount of interference from CUE $\alpha_i$ and D2D pair $j^*$ to other denied D2D pairs.
To achieve this goal the allocated power for CUE $\alpha_i$ and D2D pair $j^*$ is set to the minimum amount that guarantees their QoS requirement and are derived analytically as expressed in \eqref{EqFirstD2DAns}.
The minimum power of CUE $\alpha_i$ and D2D pair $j^*$ must be a feasible point in the sum-rate optimization problem.
\begin{equation}
\label{EqFirstD2DAns}
\left \{
\begin{matrix}
   P_{C_{\alpha_i}}^* = \frac{  (\gamma_{C_{\alpha_i}}  \gamma_{D_{j^*}}  h_{D_{j^*},B} 
+ \gamma_{C_{\alpha_i}} g_{D_{j^*}} ) ( \sigma_N^2 + \sigma_s^2)}
{ g_{C_{\alpha_i},B} g_{D_{j^*}} - \gamma_{C_{\alpha_i}}  \gamma_{D_{j^*}} h_{C_{\alpha_i},D_{j^*}}   h_{D_{j^*},B} }\\
P_{D_{j^*}}^* = \frac{  (\gamma_{C_{\alpha_i}}  \gamma_{D_{j^*}} h_{C_{\alpha_i},D_{j^*}} 
+  \gamma_{D_{j^*}} g_{C_{\alpha_i},B} ) ( \sigma_N^2 + \sigma_s^2)}
{ g_{C_{\alpha_i},B} g_{D_{j^*}} - \gamma_{C_{\alpha_i}}  \gamma_{D_{j^*}} h_{C_{\alpha_i},D_{j^*}}   h_{D_{j^*},B} }
\end{matrix}
\right .
\end{equation}
If $P_{C_{\alpha_i}}^* > P_{C,max}$ or $P_{D_{j^*}}^*>P_{D,max}$ then the power allocation problem does not have a feasible solution which means that the D2D pair $j^*$ is not an admissible D2D pair for CUE $\alpha_i$.

\begin{algorithm}[tb!]
\caption{Joint power and spectrum allocation to multiple D2D pairs in one RB underlaying a cellular network}
\label{Alg_MulD2D}
\begin{algorithmic}[1]
\STATE \textbf{Input}:   $\bm{C}$ (Set of CUEs), $\bm{D}$ (Set of D2D pairs)
\STATE \textbf{Output}: $\textbf{p}_C$ (Transmission powers of CUEs), $\textbf{p}_D$ (Transmission powers of D2D pairs), $\bm{\rho}$ (Resource sharing indicator)
\STATE \textbf{Initialization}: $\mathcal{D}=\bm{D}, $
\STATE Calculate CUE priority vector $\bm{\alpha}$ from \eqref{EqAlphaCompute}
\FORALL{$ i \in \bm{\alpha} $}
\STATE Calculate $m_j \forall j \in \mathcal{D}$ from \eqref{Eqmj} and $j^*$ from \eqref{EqmjMax}
\STATE Transmission of D2D pair $j^*$ channel gain to BS
\STATE Calculate $\mathcal{P} = (P_{\alpha_i},P_{D_{j^*}})$ from \eqref{EqFirstD2DAns}
\IF{$P_{\alpha_i}<P_{C,max}$ and $P_{D_{j^*}}<P_{D,max}$}
\STATE $\mathcal{D} = \mathcal{D} - j^*$
\REPEAT
\STATE Calculate $m'_j, \forall j \in \mathcal{D}$ from \eqref{Eqm'j} and $j^*$ from \eqref{EqMaxMin1}
\STATE Transmission of D2D pair $j^*$ channel gain to BS
\STATE Adjust $P_{C_{\alpha_i}}$ ( Section \ref{SubSec3} )
\FOR{k=1:K+1}
\STATE Adjust $P_{D_{\beta_k}}$ according to ( Section \ref{SubSec3})
\ENDFOR
\IF{$P_{C_{\alpha_i}} \leq P_{C,max}$ and $P_{D_{\beta_k}} \leq  P_{D,max} $}
\STATE $\mathcal{P} = \mathcal{P}_2$
\ELSE
\STATE Maximize $P_{C_{\alpha_i}}$ ( Section  \ref{SubSec4} )
\FOR{k=1:K+1}
\STATE Maximize $P_{D_{\beta_k}}$ ( Section  \ref{SubSec4})
\ENDFOR 
\ENDIF
\UNTIL{}
\ELSE
\STATE $P_{\alpha_i}=P_{C,max},P_{D_{j^*}}=0$, Continue
\ENDIF
\ENDFOR
\end{algorithmic}
\end{algorithm}
\subsection{Increasing The Number of Admitted D2D Pairs}
\label{SubSec3}
After selecting the first D2D pair to reuse the RB of CUE $\alpha_i$, other denied D2D pairs can also reuse the RB.
We propose to select one denied D2D pair based on a distance and power metric at each step and try to allocate the minimum possible power that guarantees QoS of UEs.
If a feasible power allocation exits, the same process will be repeated for the next selected denied D2D pair, otherwise the power of all UEs using the same RB will be maximized in order to maximize overall sum-rate.

 CUE $\alpha_i$ and D2D pairs $\beta_1, \beta_2, ..., \beta_k$ are using the same RB and D2D pair $\beta_{k+1}$ may also use the same RB.
We define $m'$ as a function of the minimum distance of CUE $\alpha_i$ and transmitter of D2D pairs $\beta_1, \beta_2, ..., \beta_k$ from the receiver of denied D2D pair $j$.
It is a power normalized distance metric because larger distances and lower transmission powers of other UEs results in less interference to the newly admitted D2D pair.
\begin{equation}
\label{Eqm'j}
m'_j = min \{ \frac{s(C_{\alpha_1},D_{j,Rx})}{P_{C_{\alpha_1}}} ,
\frac{ s(D_{\beta_1},D_{j,Rx})}{P_{D_{\beta_1}}},  ... ,\frac{s(D_{\beta_k},D_{j,Rx})}{P_{D_{\beta_k}}}  \}.
\end{equation}

The next admitted D2D pair is the one that has the largest $m'$ parameter, i.e. the one that has the highest shortest distance from other UEs,  and is the solution of the following maxmin optimization problem.
\begin{equation}
\label{EqMaxMin1}
j^* = \argmax_{j \in \mathcal{D}} {(m'_j)}.
\end{equation}

Allocation of the minimum power that guarantees the QoS of UEs in the same RB is discussed in a $K+2$ dimension space where one CUE and $K+1$ D2D pairs are using the same RB.
The channel gains of links related to the newly admitted D2D pairs are computed and transmitted to the BS.
The power allocation procedure starts from the an initial point in the $K+2$ dimension space that is defined as
$\bm{\mathcal{P}} = \{ P_{C_{\alpha_i}}, P_{D_{\beta_1}},...,P_{D_{\beta_{K+1}}}\ \}$.
where $P_{C_{\alpha_i}}$ is the transmitting power of CUE $\alpha_i$ and  $P_{D_{\beta_1}},...,P_{D_{\beta_{K}}}$ are the transmitting powers of previously admitted D2D pairs from the previous step.
The value of $,P_{D_{\beta_{K+1}}}$ is set to the maximum possible power of a D2D pair, $P_{D,max}$.
The power of CUE $\alpha_i$ is adjusted until its QoS gets equal to the its minimum required QoS and the QoS constraint become an equal active constraint.
At each step, the power of one D2D pair is adjusted on a direction that preserves the equal active constraint of previous steps.
For this purpose the increase or decrease direction of $P_{D_{\beta_k}}$ is set to be along the slope of hyperplanes related to equal active QoS constraints of other UEs in previous steps and in the same RB.
In this way a previously equal active QoS constraint remains equal active at the next steps and the power of other D2D pairs in the next steps are adjusted in a way that their QoS gets close to the minimum QoS requirement.

\subsection{Increasing the Power of CUE and admitted D2D Pairs}
\label{SubSec4}
After adding the maximum number of possible D2D pairs to the admitted D2D pair set, the optimal power must be computed which is more than their current power values and results in an increases in the sum-rate of the network.
Assuming that a maximum number of $K$ D2D pairs are reusing the RB of the CUE $\alpha_i$, an initial point $\bm{\mathcal{P}}$ is used to start the allocation of maximum feasible power to all UEs. 
This point is defined as
$ \bm{\mathcal{P}} = \{ P_{C_{\alpha_i}}, P_{D_{\beta_1}},...,P_{D_{\beta_{K}}} \} $
where each the power of each UE is the minimum QoS guaranteeing power from the previous step.
The maximization of power is performed at $K+1$ steps.
At the first step, the power of CUE $\alpha_i$ is adjusted until the QoS constraint related to another UE become equal to its minimum QoS requirement.
Then this procedure is repeated for all other UEs that are using the same RB of CUE $\alpha_i$ at next steps.
Similar to previous section at each step the power of one D2D pair in increased on a direction that preserves the equal active constraint of previous steps.
The increase direction of $P_{D_{\beta_k}}$ is set to be along the slope of hyperplanes related to QoS constraints that are equal to their minimum QoS requirement.
So the maximum feasible transmitting power of all UEs is reached while their QoS is also guaranteed. 
At each step of the algorithm, the QoS constraint of a UE with a QoS equal to its minimum requirement is preserved  and the power of other UEs are increased.
Hence overall sum-rate is maximized.
The overall resource allocation algorithm is presented at table \ref{Alg_MulD2D}.

\section{Complexity and Signaling Overhead Analysis}
\label{SecComp}
Considering the assumption of at most one D2D pair at each RB, there will be $M+1$ possible matching state for each CUE.
But in our system model  there will be $2^M$ possible matching states for each CUE.
The total number of matching states for optimal matching $N_{\rm{states, optimal}}$ will be equal to 
\begin{equation}
\label{NStatesOptimal}
N_{\rm{states,optimal}} = N  \sum_{k=0}^{k=M} {\binom{M}{k}} = N 2^M
\end{equation}
Based on our suboptimal and low complexity proposed matching procedure, there will be $N$ steps related to all CUEs and at most a mean number of $M/N$ D2D pairs matched to the CUE at each step. Thus the complexity of our proposed algorithm will be $N_{\rm{states,proposed}} \approx N (M/N) = M$.
In our system model the number of channel gains related to the optimal resource allocation is
\begin{equation}
\label{NSignalingOptimal}
N_{\rm{signaling, optimal}} = N(M+1) +2M + M(M-1)
\end{equation}
In our proposed method, the channel gain between UEs that are using the same RB is just needed.
Assuming a uniform load distribution, a maximum number of $M/N$ D2D pairs are matched to one specific CUE and the signaling overhead $N_{\rm{signaling, proposed}}$ will be equal to 
\begin{equation}
\label{NSignalingProposed}
N_{\rm{signaling, proposed}} \approx N(M+1) +2M + M(M/N-1) 
\end{equation}

\section{Numerical Results}
\label{SecNumeric}
A single cell network with uniformly distributed CUEs is considered.
Transmitter and receiver of each D2D pair are also uniformly distributed in a cluster with radius $r$ where different D2D pairs are located in different clusters and each cluster is also uniformly distributed in the cell.
QoS requirements, maximum transmit power of UEs, and the number of UEs are described at table I.

\begin{table}[t]
\label{TSimParams}
\begin{center}
\caption{Simulation Parameters} \label{T2}
\vspace*{-0.2cm}
\def\arraystretch{1.2}
\begin{tabular}{|p{0.9in}|p{1in}|p{1in}|} \hline
\multicolumn{2}{|p{1.8in}|}{\textbf{Parameter}}& \textbf{Value}  \\ \hline
\multicolumn{2}{|p{1.8in}|}{Physical link type} & {Uplink} \\ \hline
\multicolumn{2}{|p{1.8in}|}{Cell Radius} & {400, 600 m} \\ \hline
\multicolumn{2}{|p{1.8in}|}{Noise power ($\sigma^2_N$)} & -114 dBm \\ \hline
\multicolumn{2}{|p{1.8in}|}{Path loss exponent ($\alpha$)} & -3.5 \\ \hline
\multicolumn{2}{|p{1.8in}|}{Maximum D2D Tx power ($P_{D,max}$)} & 18 dBm\\ \hline
\multicolumn{2}{|p{1.8in}|}{Maximum CUE Tx power ($P_{C,max}$)} & 24 dBm\\ \hline
\multicolumn{2}{|p{1.8in}|}{CUE and D2D minimum QoS requirements ($\gamma_{C_{i,min}}, \gamma_{D_{j,min}}$)} & Uniformly distributed in [5,20] dB\\ \hline
\multicolumn{2}{|p{1.8in}|}{D2D cluster radius} & 10,15, ..., 40 (m)\\ \hline
\multicolumn{2}{|p{1.8in}|}{Number of CUEs (N)} & 5,10 \\ \hline
\multicolumn{2}{|p{1.8in}|}{Number of D2D pairs (M)} & 5N \\ \hline
\end{tabular}
\end{center}
\vspace*{-0.6cm}
\end{table}

\begin{figure}[h!]
  \centering
  \includegraphics[width=\linewidth]{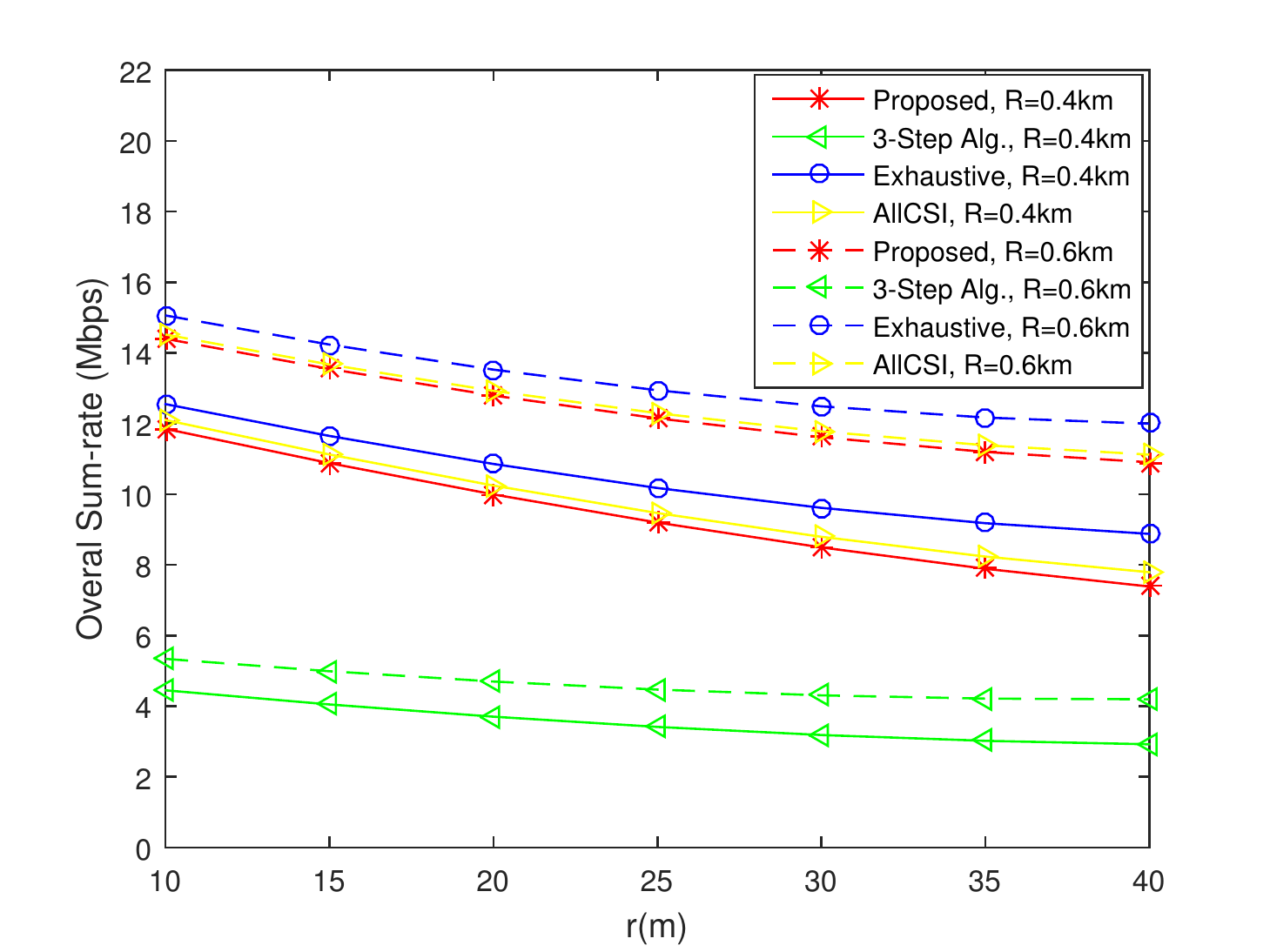}
  \caption{Sum-rate for different D2D cluster radius ($R=0.4, 0.6 km$, $N=5$).}
  \label{Fig2}
\end{figure}
\begin{figure}[h!]
  \centering
  \includegraphics[width=\linewidth]{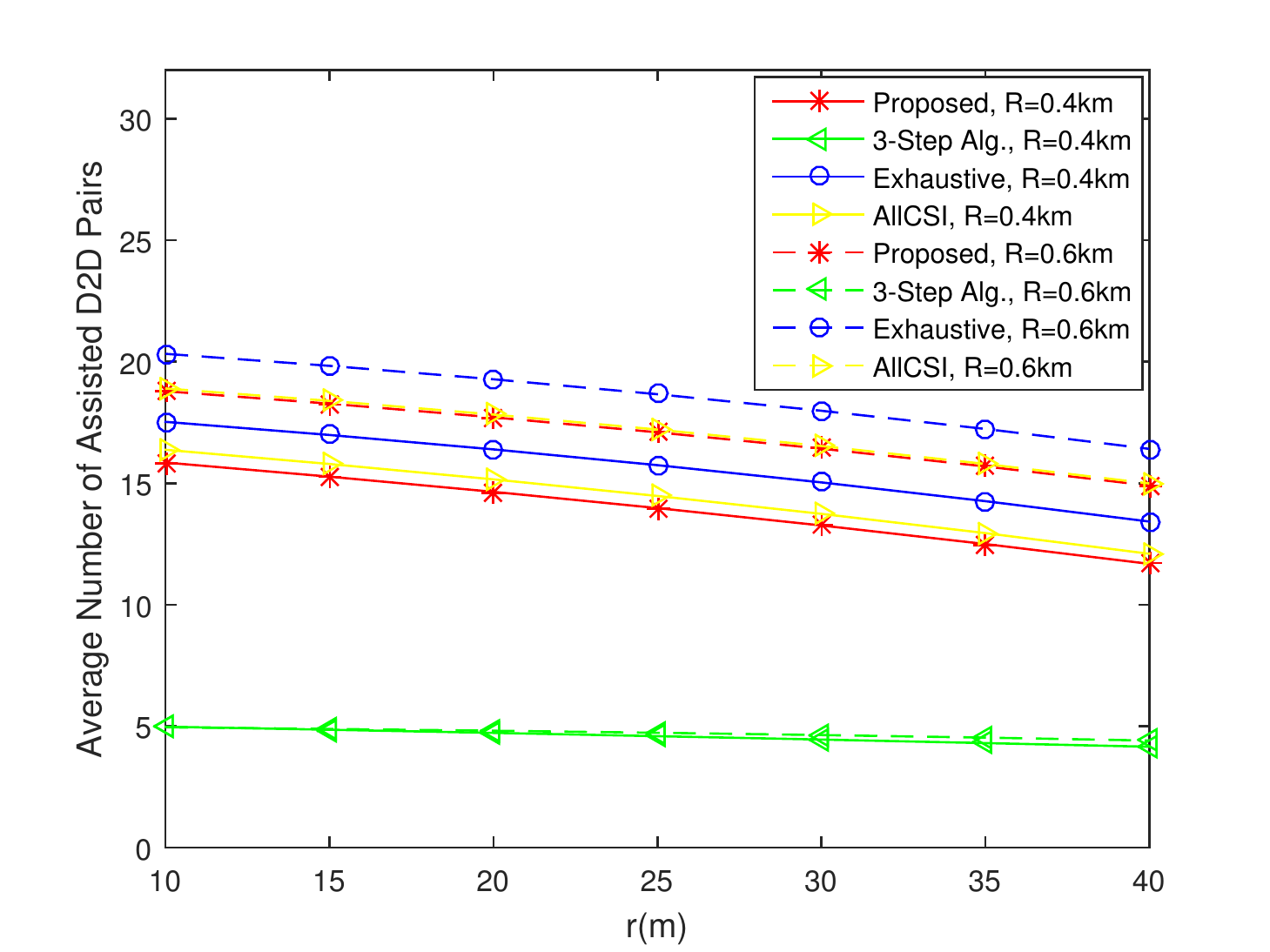}
  \caption{Average number of assisted D2D pairs ($R=0.4, 0.6 km$ and $N=5$).}
  \label{Fig3}
\end{figure}
\begin{figure}[h!]
  \centering
  \includegraphics[width=\linewidth]{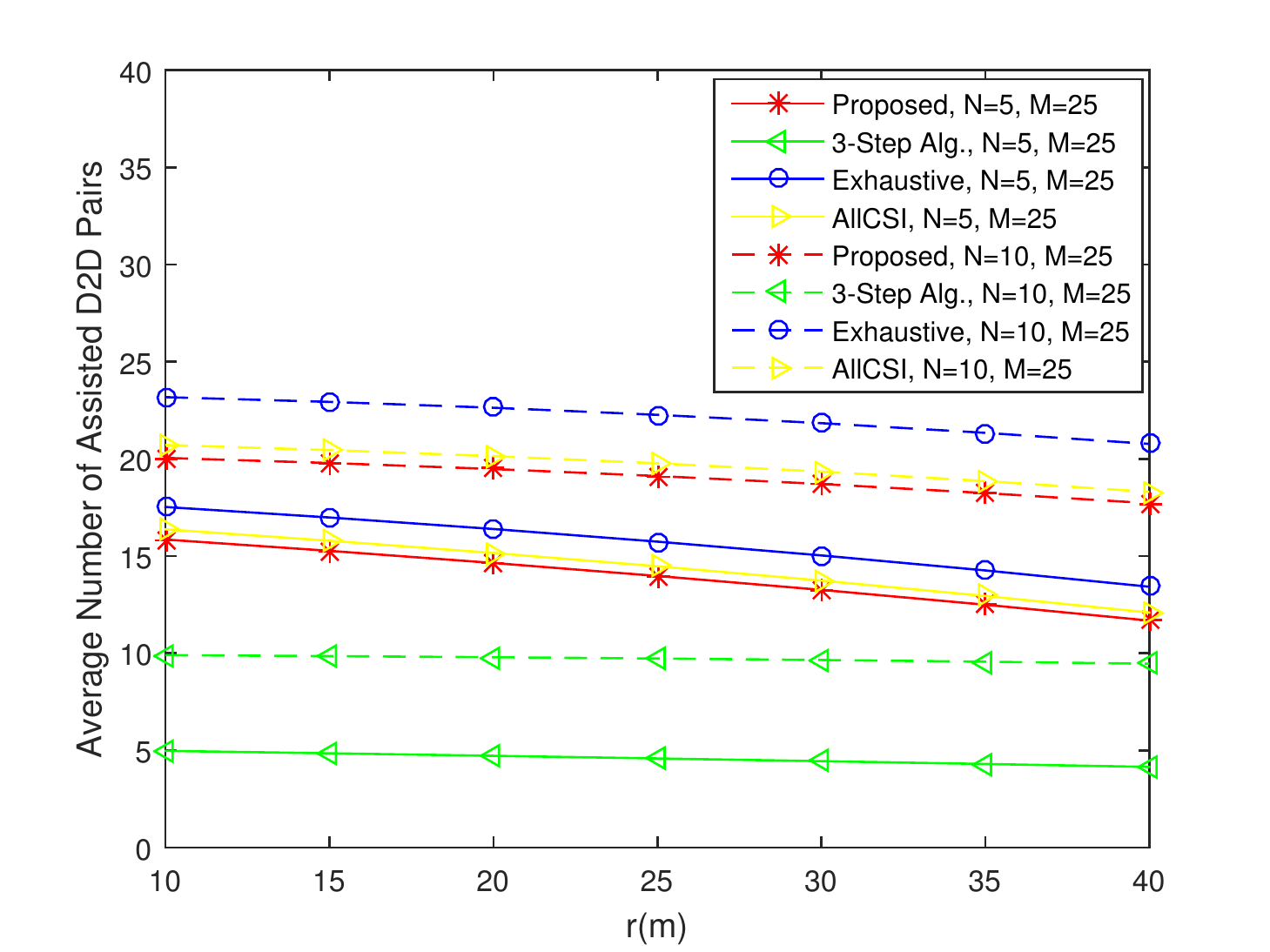}
  \caption{Average number of assisted D2D pairs  ( $N=5, 10$, $R=0.4 km$, and $M=25$).}
  \label{Fig4}
\end{figure}

Overall sum-rate and the number of assisted D2D pairs are used as performance evaluation metrics.
We compare our scheme with 3-step, AllCSI, and exhaustive search algorithms.
The 3-step algorithm is proposed in \cite{R003} and assumes that the RB of each CUE can be shared with at most one D2D pair. This method assumes known channel gains at the BS and computes overall sum-rate for all matching states.
The AllCSI algorithm  is presented at \cite{RKhal} and assumes that the channel gains of all links are known at the BS.
This method has a huge signaling overhead equal to \eqref{NSignalingOptimal} and uses a reduced complexity matching algorithm that selects the next admitted D2D pair based on minimum interference link gain.
The exhaustive search algorithm is a generalized approach of the 3-step algorithm that is applied to a large number of CUE-D2D matching states.
The authors of \cite{R003} have proposed the 3-step algorithm and we have generalized their approach to consider multiple D2D pairs in one RB.
This exhaustive algorithm has a large number of matching states equal to \eqref{NStatesOptimal} and a huge signaling overhead equal to \eqref{NSignalingOptimal} and can be considered as an optimal solution to the problem.

Figure \ref{Fig2} compares the overall sum-rate performance of the four algorithms for different cell radii.
The 3-step algorithm has the lowest overall sum-rate since a maximum of one D2D pair can reuse the RB of each CUE.
AllCSI and the proposed algorithms have lower performances compared to optimal algorithm but with less complex matching procedures.
The optimal algorithm has an exponential matching complexity while the matching complexity of the proposed algorithm increases linearly with $M$.
The performance of AllCSI algorithm is slightly higher than the proposed algorithm due to its huge signaling overhead.
The signaling overhead of our proposed method is $N$ time less than AllCSI algorithm.
The performance gap between AllCSI and the proposed algorithm is considerably reduced when the cell radius is increased.

Figure \ref{Fig3} discusses the number of admitted D2D pairs for different cell radii.
The number of admitted D2D pairs in the 3-step algorithm is almost equal to the number of CUEs, while the number of admitted D2D pairs in AllCSI and the proposed method is almost three times more than the number of CUEs.
Due to previously mentioned reasons related to complexity and signaling overhead, the performance of exhaustive search algorithm is higher than the two other algorithms.
By increasing the radius of the cell, more D2D pairs can be admitted.

Figure \ref{Fig4} considers a constant number of D2D pairs with different number of CUEs.
It can be seen that increasing the number of CUEs results in an increase in the number of RBs.
So better interference management can be done and more D2D pairs can be admitted.

It is shown that the proposed algorithm significantly increases the overall sum-rate compared to the situation where at most one D2D pair is admitted for each RB and is slightly lower than the performance of optimal algorithm for multiple D2D pairs in one RB.
The matching complexity of the proposed algorithm increases linearly respect to $M$ while it grows exponentially in the optimal algorithm.
The proposed algorithm also has smaller signaling overhead while optimal and AllCSI algorithms have a huge signaling overhead.

\section{Conclusion and Future Work}
\label{SecConclusion}
In this paper we have investigated joint power and spectrum allocation for D2D communications in a centralized manner for sharing uplink resources in a fully loaded cellular network with multiple D2D pairs reusing the same RB in order to maximize overall sum-rate.
The optimal solution of this problem has a high complexity CUE-D2D matching and signaling overhead.
We propose to allocate resource to each CUE at one step where the maximum possible number of denied D2D pairs are added to the admitted D2D pair set of that CUE.
At each step of adding denied D2D pairs to the admitted D2D pair set, the power of all UEs are minimized with respect to the QoS constraints using a QoS hyperplane slope based approach.
The power of all UEs in the same RB are maximized when no more denied D2D pair can be added.
Our proposed suboptimal algorithm dramatically reduces the matching complexity and signaling overhead.
In future research, we will consider the problem with channel uncertainty.
Also deriving a close form for power allocation related to sections \ref{SubSec2} and \ref{SubSec3} should be investigated.

\bibliographystyle{IEEEtran}
\bibliography{Test}

\end{document}